\documentclass[pra,twocolumn,showpacs,preprintnumbers,amsmath,amssymb]{revtex4}
\usepackage{graphicx}% Include figure files
\usepackage{dcolumn}% Align table columns on decimal point
\usepackage{bm}% bold math
\usepackage{latexsym,epsfig}
\usepackage{textcomp}

\newcommand{\beq}{\begin{equation}}
\newcommand{\eeq}{\end{equation}}
\newcommand{\bea}{\begin{eqnarray}}
\newcommand{\eea}{\end{eqnarray}}
\newcommand{\ben}{\begin{eqnarray*}}
\newcommand{\een}{\end{eqnarray*}}
\newcommand{\be}{\begin{enumerate}}
\newcommand{\ee}{\end{enumerate}}
\newcommand{\bfig}{\begin{figure}}
\newcommand{\efig}{\end{figure}}

\newcommand{\ba}{\begin{align}}
\newcommand{\ea}{\end{align}}

\begin{document}
\preprint{Version 4.6}
\title{Quantum Phases of Ultracold Bosonic Atoms in a One Dimensional Optical Superlattice}
\author{Arya Dhar}
\email{arya@iiap.res.in}
\affiliation{Indian Institute of Astrophysics, II Block, Koramangala, Bangalore-560 034, India}
\author{Tapan Mishra}
\email{tapan@physics.georgetown.edu} 
\affiliation{Department of Physics, Georgetown University,
Washington, DC 20057, USA}
\author{Ramesh V. Pai}
\email{rvpai@unigoa.ac.in} \affiliation{ Department of Physics, Goa
University, Taleigao Plateau, Goa 403 206, India. }
\author{B. P. Das}
\email{das@iiap.res.in} \affiliation{Indian Institute of
Astrophysics, II Block, Kormangala, Bangalore, 560 034, India.}

\date{\today}

\begin{abstract}
We analyze various quantum phases of ultracold bosonic atoms in a periodic one dimensional
optical superlattice. Our studies have been performed using the
finite size density matrix renormalization group
(FS-DMRG) method in the framework of the Bose-Hubbard model. Calculations have been carried out for a
wide range of densities and the energy shifts due to the
superlattice potential. At commensurate fillings, we find the Mott
insulator and the superfluid phases as well as Mott insulators 
induced by the superlattice. At a particular incommensurate density, the system is found to be in the
superfluid phase coexisting with density oscillations for a certain range of parameters of the system.
\end{abstract}



\maketitle

\section{INTRODUCTION}

%{\it  (Here you can give some text on superlattice. Please go to Inguscio's home page
%and see what he has done on  superlattice. Also read Marcos rigol's paper to write something about superlattice research.
%We also should mention about the supersolid in superlattice with
%dipolar atoms which I had sent to you recently).}
%{\it while writing about superlattice you can give figure of superlattice.}

%Studies of quantum phase transitions are currently a very active
%research area of great interest~\cite{rmp}.  The physics of
%ultracold atoms entered the areas of strongly correlated systems
%with the seminal 1998 paper by Jaksch et al ~\cite{jakschnature}
%which predicted the superfluid(SF) - Mott insulator(MI) transition in
%cold atoms loaded in an optical lattice. The pioneering observation
%of the SF-MI transition in an optical lattice using cold bosonic
%atoms ~\cite{bloch} marked the beginning of age of the experimental
%studies of strongly correlated systems with ultracold atoms. It also
%highlighted the possibility for detailed investigations of strongly
%correlated quantum systems in such perfectly controllable systems.

In a seminal paper in 1998, Jaksch \textit{et al.} ~\cite{jaksch} extended the
work of Fisher \textit{et al.}~\cite{fisher} and predicted the superfluid
(SF) - Mott insulator (MI) transition in ultracold bosonic atoms in
an optical lattice. The observation of this transition
in an optical lattice marked
the beginning of the experimental studies of quantum phase
transitions in ultracold atoms arising from strong
correlations~\cite{greiner}. It also highlighted the possibility for
detailed investigations of various kinds of strongly correlated quantum
systems in optical lattices where the different parameters of
these systems can be exquisitely
controlled~\cite{lewenstein,bloch,dalibard}.
\par

In this paper, we analyse the various quantum phases exhibited by ultracold bosonic atoms in a periodic
one dimensional optical superlattice using the finite-size density matrix
 renormalisation group (FS-DMRG) method for certain parameters of the system. Previously, it has been shown using
the exact diagonalization, quantum Monte Carlo and the mean-field
decoupling approximation methods that ultracold bosonic atoms in optical superlattices exhibit
different phases, with various charge-density-wave orders
apart from the usual SF and MI
phases~\cite{rousseau,rothpra,shuchen}. In the presence of a disordered
and quasiperiodic potential, it has been shown that the system
exhibits apart from SF and MI phases, the quasi-Bose glass and the
incommensurate charge-density wave phases~\cite{roth,giamarchi}. In
our present work, we vary the densities of the bosonic atoms from
commensurate to incommensurate values for a relatively large value of the onsite interaction.
The purpose of this investigation is to predict quantum phases that arise due to
the influence of the superlattice in the parameter space that we have considered.

\par
The paper is arranged in the following manner. The remainder of this section describes the model for bosonic atoms in a one dimensional optical superlattice that we have chosen in our work. In Section II, the FS-DMRG method and the quantities that we have calculated are briefly discussed. The results are presented in Section III and our conclusions are summarized in Section IV.

\begin{figure}[h]
\centering \psfig{file=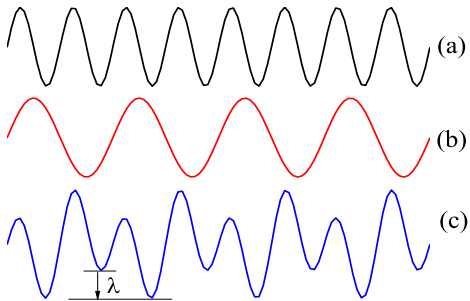,width=7cm,height=5cm}
\caption{(Color online) Superlattice (c) formed by the superposition
of two optical lattices (a) and (b), frequency of one being the
double of the other. This results in an energy shift of $\lambda$ for
alternate sites.} \label{fig:fig1}
\end{figure}

An optical superlattice is formed by the superposition of two optical
lattices of different frequencies [Fig.~(\ref{fig:fig1})]. Ultracold
atoms in an optical superlattice can be
described by a modified Bose-Hubbard model where the superlattice potential
is explicitly included.
\begin{equation}
 H=-t\sum_{\langle{i,j}\rangle}{(\hat{a}_{i}^{\dagger}\hat{a}_{j}+h.c)}+\frac{U}{2}\sum_{i}{\hat{n}_{i}(\hat{n}_{i}-1)}+\sum_{i}{\lambda_{i}\hat{n}_{i}}
 \label{eq:ham}
\end{equation}
In the above equation, $\lambda_{i}$ denotes the shift in the energy
levels for each site due to the superlattice potential. For our work,
we have considered an optical superlattice with the frequency of one optical
 lattice being double that of the other. Hence, the unit cell comprises of two adjacent lattice sites
 as depicted in Fig.~(\ref{fig:fig1}). As a result,
$\lambda_{i}=\lambda$; $\forall$ $i=$ odd integers and $\lambda_{i}=0$;
$\forall$ $i=$ even integers. In the first term of
Eq.~(\ref{eq:ham}), $\langle{i,j}\rangle$ denotes pair of nearest
neighbor sites $i$ and $j$, $t$ denotes the hopping amplitude
between them with $\hat{a}_{i}^{\dagger}$($\hat{a}_{i}$) being the
creation (annihilation) operator which creates (destroys) an atom at
site $i$. In the second term of Eq.~(\ref{eq:ham}), $U$ represents
the on-site inter-atomic interaction and
$\hat{n}_{i}=\hat{a}_{i}^{\dagger}\hat{a}_{i}$ is the number
operator. The Hamiltonian is re-scaled in units of hopping amplitude
$t$ by setting $t=1$ so that all the quantities becomes
dimensionless.

%Here $\langle{i,j}\rangle$ denotes pair of nearest neighbor sites
%$i$ and $j$, $t$ denotes the  hopping amplitude between them, $U$
%represents the on-site inter-atomic interaction,
%$\hat{a}_{i}^{\dagger}$($\hat{a}_{i}$) is the creation
%(annihilation) operator which creates (destroys) an atom at site $i$
%and $\hat{n}_{i}$ is the number operator. The Hamiltonian is
%re-scaled in units of hopping amplitude $t$ by setting $t=1$ so that
%all the quantities becomes dimensionless.

In the absence of the superlattice potential, the model given in
Eq.~(\ref{eq:ham}) reduces  to the well studied Bose-Hubbard model which
exhibits a transition from the gapless, compressible superfluid (SF)
phase to the gapped, incompressible Mott insulator (MI) phase at integer
densities~\cite{fisher}. However, for non-integer densities, the
system remains in the superfluid phase.

The effect of the superlattice potential is to break the translational
symmetry of the system. This leads to a change in the local (on site)
chemical potential. In the present case, the local chemical
potential changes at alternate sites in a periodic manner. The
SF-MI transition has been predicted for such a
system at interger fillings~\cite{shuchen}. Apart from the SF phase,
MI phases with various
fillings at integer and half-integer densities have been predicted depending on the
relative values of the different parameters of the system ~\cite{shuchen,roth}. The
MI phase is characterized by a fixed number of atoms in a particular
site or more
 specifically in a unit cell, and it occurs due to the strong onsite interaction between the atoms.
In the two-period superlattice that we have considered, it is possible to have a new
class of insulators in which alternate lattice sites occupy a fixed number of atoms. However unlike the
MI phase which is due to strong interatomic interactions, these insulators arise because of the
superlattice potential. We therefore refer to these insulators as superlattice
induced Mott insulators (SLMI). If odd sites are occupied by one atom
and none in the even ones, then we call it as the SLMI-I phase whereas, if
odd sites are occupied by 2 atoms with none in the even, then it is
called the SLMI-II phase.

\section{METHOD OF CALCULATION}
To obtain the ground state wave function and the energy eigenvalues
corresponding to  model (\ref{eq:ham}) for a system of $N$ bosons in
a lattice of length $L$, with on-site interactions and tunnelling as
well as the energy shift in alternate lattice sites due to the
superlattice potential, we use the FS-DMRG method with open boundary
conditions ~\cite{white,schollwock}. This method is best suited for
one-dimensional problems and has been successfully used to study the
Bose-Hubbard model ~\cite{kuhner,pai,monien,schollwock,urba}. In our
computations, we have considered four bosonic states per lattice site
and the weights of the states neglected in the density matrix formed
from the left or the right blocks are less than $10^{-6}$
~\cite{pai}. In order to improve the convergence, at the end of each
DMRG step, we use the finite-size sweeping procedure given in
~\cite{white,pai}. To obtain the ground state wave-function
$|\varPsi_{LN}\rangle$, and the corresponding energy $E_{L}(N)$ for
densities ranging from $0.24$ to $1.25$, we start with $4$ sites and
$4$ atoms in the FS-DMRG and increase both of them by two at each
iteration, till we have $24$ atoms. Then we do not increase the
number of atoms, but increase the number of lattice sites to
$100$ by adding two sites in each DMRG iteration. After each step of
the iteration, sweeping was done from left to right and also right to left
across the entire lattice. This
process is continued till the energy converges. The
superlattice potential breaks the symmetry between the system
and the environment. Therefore it is necessary to perform the calculations on the
system and the environment blocks separately at each DMRG iteration. Once the length $L=100$ is reached, we keep it fixed and then increase the number of atoms by adding one at a
time and perform a complete DMRG sweep for the convergence of the
energy. This iterative procedure is continued till the number of bosons is
equal to $125$. By this process we obtain the wavefunctions
and the corresponding energies of model(~\ref{eq:ham}) for densities
ranging from $0.24$ to $1.25$. We have carried out the calculations for a
range of values of $\lambda$ starting from $0$ to $15$, with $t$ and
$U$ fixed at $1.0$ and $10.0$ respectively.

Using the ground state wave-function
$|\varPsi_{LN}\rangle$, and the corresponding energy $E_{L}(N)$, the
following quantities are calculated to identify the various phases
exhibited by the system. Firstly the chemical potential of a system
for the density $\rho=N/L$ is defined as follows:
\begin{equation}
 \mu=\frac{\delta{E}_{L}(N)}{\delta{N}}
 \label{eq:mu}
\end{equation}
The gapped and gapless phases are distinguished from the behavior of
$\rho$ as a function of $\mu$~\cite{ramanan}. To get information
about the on-site density distribution in various phases exhibited
by the system, we calculate the on-site density,
$\langle{n_{i}}\rangle$, defined as,
\begin{equation}
 \langle{n_{i}}\rangle=\langle{\varPsi_{LN}|\hat{n}_{i}|\varPsi_{LN}}\rangle.
\end{equation}
The momentum distribution $n(k)$ and the structure function $S(k)$
are defined as the Fourier transform of the single particle density matrix
$\langle{\hat{a}^{\dag}_{p}\hat{a}_{q}}\rangle=\langle{\varPsi_{LN}|\hat{a}^\dag_{p}\hat{a}_q|\varPsi_{LN}}\rangle$
and the density-density correlation function
$\langle{\hat{n}_{p}\hat{n}_{q}}\rangle=\langle{\varPsi_{LN}|\hat{n}_{p}\hat{n}_q|\varPsi_{LN}}\rangle$,
respectively:
\begin{equation}
 n(k)=\frac{1}{L}\sum_{p,q}{e^{ik(p-q)}\langle{\hat{a}_{p}^{\dagger}\hat{a}_{q}}\rangle}
\label{eq:momdistri}
\end{equation}

\begin{equation}
 S(k)=\frac{1}{L}\sum_{p,q}{e^{ik(p-q)}\langle{\hat{n}_{p}\hat{n}_{q}}\rangle}.
\label{eq:strucfunc}
\end{equation}

The momentum distribution, $n(k)$ will indicate the presence of SF phase in the system. The structure function will give information about the presence of any density wave order in the system, and also the effect of superlattice potential on the Brillouin zone boundaries.

To get an idea of the various gapped and gapless phases in the
system, we plot the density, $\rho$ vs the average chemical
potential, $\mu$. However to draw the phase diagram, we need
accurate values of the chemical potential, and hence we calculate
the chemical potential (both $\mu^{+}$ and $\mu^{-}$) in the
thermodynamic limit $L \rightarrow \infty$. $\mu^{+}_L$ is defined
as the difference in the energy of the system of length $L$ if one
atom is added to the system, whereas $\mu^{-}_L$ is the energy cost
when one atom is removed from the system, i.e.,
$\mu^+_L=E_L(N+1)-E_L(N)$ and $\mu^-_L=E_L(N)-E_L(N-1)$. We plot
these chemical potential values versus the reciprocal of the length ($1/L$), and then
extrapolate to length tending to infinity ($1/L \rightarrow 0$) to
get the values of the chemical potential in the thermodynamic limit. We plot
these chemical potential values for both density half and one for
different superlattice potential $\lambda$ to get the phase diagram.
%From the behavior of the
%above physical quantities mentioned in Eqs. ~(\ref{eq:mu})
%-~(\ref{eq:strucfunc}), we obtain the various phases exhibited by the model
%(~\ref{eq:ham}).

%While evaluating the energyeigen values and the wavefunctions by FS-DMRG, the length of the system is 4 in the beginning and increased at every
%iteration. We find the energy values at each such length, with number of atoms being equal to length, one more than and one less than length %respectively to calculate $\mu^{+}$ and $\mu^{-}$ for density one. For density equal to half, appropriate changes are made in the initial number of %atoms, and also the increment in the number of atoms at each iteration. We plot these chemical potential values versus inverse of length, and then %extrapolate to length tending to infinity (1/L=0). Thus we get chemical potential values in the thermodynamic limit. We plot these chemical potential %values for both density half and one versus the superlattice potential, $\lambda$, to get the phase diagram.

%To obtain densities ranging from 0.24 to 1.25, we start with 4 sites
%and 4 atoms in the FS-DMRG and increase both of them by two at each
%iteration, till the number of atoms reach 24. Then we stop
%increasing the number of atoms, and increase the length to 100 by
%adding two sites in each DMRG iteration. After each step of the
%iteration, sweeping was done from left to right and vice-versa
%across the entire lattice so as to optimize the choice of the
%hamiltonian matrix, both for the system and the environment.

\section{RESULTS AND DISCUSSION}

Before we discuss the details of our results, we first summarize the
main features of the phase diagram. It is well established that
bosons in a normal optical lattice, described by the Bose-Hubbard
model~\cite{fisher}, exhibit the superfluid to Mott insulator quantum
phase transition at some critical value of $U$ for integer densities. (For example, for
$\rho=1$ this transition occurs at the critical on-site interaction
$U_C\sim 3.4$~\cite{kuhner,pai} in one-dimension). For non-integer
densities, only the superfluid phase exists in the ground state. The
Mott insulator phase has a finite gap and is incompressible. On the
other hand, the superfluid phase is gapless and compressible. When
we include the superlattice potential, the energy levels at each
site is shifted by $\lambda_i$ and this leads to the additional incompressible, gapped SLMI phase. In model~(\ref{eq:ham}), the SLMI phase occurs
at commensurate densities, i.e., when $\rho$ is equal to an integer or an
half-integer, because the superlattice we have considered in our
calculations has a periodicity of two lattice sites. For $\rho=1/2$, we observe a SF to SLMI-I phase
transition as the strength of the superlattice potential is
increased. However, for $\rho=1$ we get the MI and the SLMI-II phases along
with a SF phase sandwiched between them. Such a transition at $\rho=1.0$ had already been predicted in an optical superlattice~\cite{rousseau} using a quantum Monte Carlo approach. However this SF phase at $\rho=1$ is not the usual SF phase
which arises due to a large value of $t$/$U$. In fact, in our
present case, the ratio is very small ($t/U\sim 0.1$). This SF phase
is a result of the competition between the superlattice potential,
$\lambda$ and the on-site repulsive interaction potential, $U$. Moreover, because of the superlattice potential, this SF phase will have finite density modulations.

\begin{figure}[htbp]
\centering \psfig{file=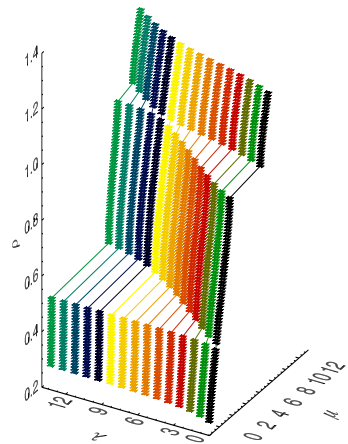, height=3.5in,
width=3.5in,angle=0} \caption{(Color online) Density $\rho$ is
plotted against the chemical potential for various values of
$\lambda$ at a fixed $U=10$ and $t=1$.} \label{fig:fig2}
\end{figure}

We first discuss the results for commensurate densities and then
for incommensurate densities. In the Fig.~(\ref{fig:fig2}) we
plot $\rho$ versus $\mu$, defined by Eq.~(\ref{eq:mu}), for a fixed
value of $U=10$, but $\lambda$ varying from $0$ to $15$.  The gapped
phases show up in this type of plots as plateaus with the gap $G_L$
equal to the width of the plateau, i.e., $G_L=\mu^+_L-\mu^-_L$,
where $\mu^+_L$ and $\mu^-_L$ are the values of the chemical
potential at the upper and lower knees of the plateau, respectively.
%\vspace{0.5cm}
%\begin{figure}[htbp]
% \centering
%\epsfig{file=rho_mu_plot1.eps, height=8cm, width=8cm, angle=0}
%\caption{The chemical potential, $\mu$ is plotted against density
%$\rho$ for $\lambda=0.2, 2.2, 4.2, 6.2$, and $U=10.0$}
%\label{fig:fig3}
%\end{figure}

\begin{figure}[htbp]
 \centering
 \psfig{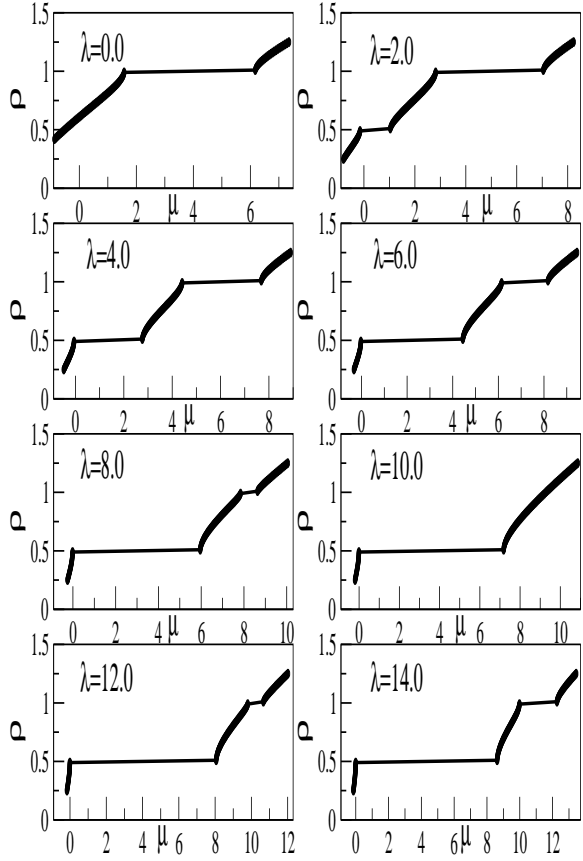}
\caption{The chemical potential, $\mu$ is plotted against density
$\rho$ for various values of $\lambda$, and $U=10$.} \label{fig:fig3}
\end{figure}
\par
In Fig.~\ref{fig:fig2} we see for $\lambda=0$, there exists only one plateau at $\rho=1.0$ correponding to the
 MI phase, which is expected
since the value of $U$ is very large compared to $t$~\cite{pai}. This MI phase survives for
small values of $\lambda$. As the strength of $\lambda$ increases,
two interesting features appear in the plots: (i) a new plateau
appears at $\rho=1/2$ and (ii) the width of the plateau at $\rho=1$
decreases. This can be clearly seen in the plots of $\rho$ versus
$\mu$ for various $\lambda$ values as shown in Fig.~(\ref{fig:fig3}). For $\lambda=0.2$, the gapped phase exists
only at $\rho=1$ [Fig.~(\ref{fig:fig3})] and is gapless at
$\rho=1/2$. As $\lambda$ is increased beyond $0.8$, two gapped
phases appear, at $\rho=1/2$ and $\rho=1$, respectively. The gapped
phase at $\rho=1/2$ occurs due to the transition from the SF to the SLMI-I phase
with one atom per unit cell, occupying alternate sites.
%\vspace{0.7cm}
%\begin{figure}[htbp]
% \centering
%\epsfig{file=rho_mu_plot2.eps, height=8cm, width=8cm, angle=0}
%\caption{The chemical potential, $\mu$ is plotted against density
%$\rho$ for $\lambda=8.2, 10.2, 12.2, 14.2$, and $U=10.0$}
%\label{fig:fig4}
%\end{figure}

For $\lambda$ values between $0.8$ and $9.6$, the system exhibits two
gapped regions, one at $\rho=1/2$ and the other at $\rho=1$.
The gap at $\rho=1/2$ increases steadily as $\lambda$ increases and remains finite. On the other hand, for
values of $\lambda > 0.0$, the gap at $\rho=1$ decreases
continuously and ultimately vanishes at around $\lambda=9.6$.
 This kind of behavior for $\rho=1$ is due to the competition between the
superlattice potential, $\lambda$ and on-site repulsive interaction, $U$. The
on-site interaction $U$ tries to impose the MI phase in the system
whereas $\lambda$ tends to introduce the SLMI-II phase. As long as
$U$ is greater than $\lambda$, the MI phase is energetically favorable
for $\rho=1$ resulting in a finite gap. As $\lambda$ becomes
comparable to the value of $U$ (here in this calculation we have
fixed $U=10$), neither the MI phase, with one boson at every
site nor the SLMI-II phase, with two bosons at every alternate site is
energetically favorable. This leads to vanishing of the
energy gap and the system becomes a SF.

When $\lambda$ is larger than the value of $U$, the superlattice potential dominates over the on-site interaction. As a result,
the gap at $\rho=1$ opens up again, which can be seen clearly from
Fig.~(\ref{fig:fig2}) and~(\ref{fig:fig3}). This gap is due to the system being in the SLMI-II phase. The regions
other than commensurate densities, i.e. $\rho=1/2$ and $1$ remain in the gapless SF phase for all values of $\lambda$.
%~\cite{tapan}.
%\vspace{2.0cm}
\begin{figure}[tbp]
 \centering
\psfig{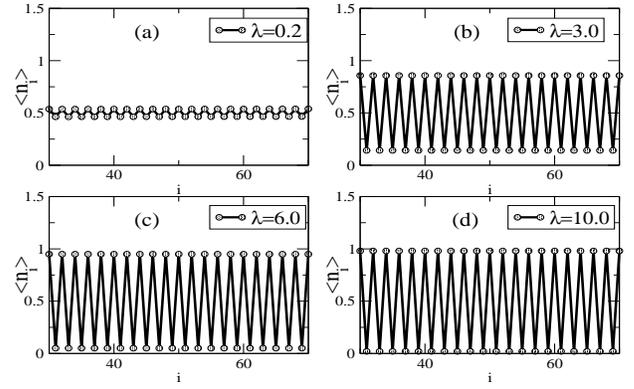}
\caption{On-site number density plotted against lattice site index at density, $\rho=0.5$.}
\label{fig:fig4}
\end{figure}
%\vspace{2.0cm}
\begin{figure}[bp]
 \centering
\psfig{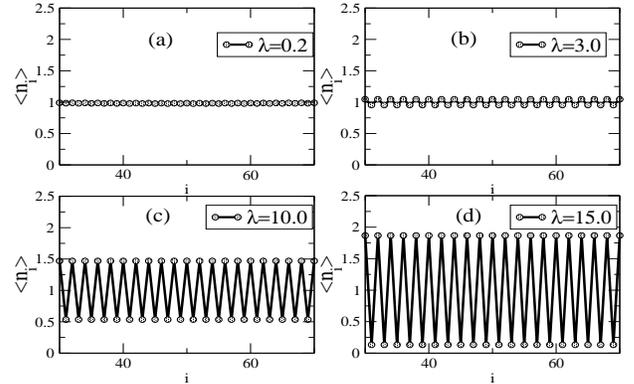}
\caption{On-site number density plotted against lattice site index at density, $\rho=1.0$.}
\label{fig:fig5}
\end{figure}

To confirm the above findings concerning the existence of the MI and the
SLMI phases, we plot the on-site average number density
$\langle{\hat{n}_i}\rangle$ with respect to the site index $i$ as
shown in Fig.~(\ref{fig:fig4}) and ~(\ref{fig:fig5}), for $\rho=1/2$ and $1$, respectively. For values of $\lambda$ less than $0.8$, there are no oscillations at density equal to half.
However, as the value of $\lambda$ is increased, the density
oscillations begin to set in the system, and at higher values of
$\lambda$, a clear $\{1~ 0~ 1~ 0~ 1~ 0~ \cdots \}$ occupancy
configuration can be seen confirming the presence of SLMI-I phase [Fig.~(\ref{fig:fig4})].
At $\rho=1$, for low values of $\lambda$, the plot [Fig.~(\ref{fig:fig5})] shows a constant number density equal to $1.0$ implying the MI phase.
As $\lambda$ is increased, but less than $9.6$, the superlattice potential induces some oscillations, but the system is still in the MI phase.
 Even at $\lambda=10.0$, the oscillations vary from
$1.5$ (maximum) to $0.5$ (minimum), significantly different from the SLMI-II
configuration of $\{2~ 0~ 2~ 0~ 2~ 0~ \cdots \}$. Once $\lambda$
becomes much larger than $U$, the SLMI-II configuration is observed. For example, at
values of $\lambda=14$, the number occupancy in the sites look very
similar to the SLMI-II configuration as can be seen from figure
[Fig.~(\ref{fig:fig5})].
%\vspace{2.0cm}
\begin{figure}[tbp]
\centering
 \psfig{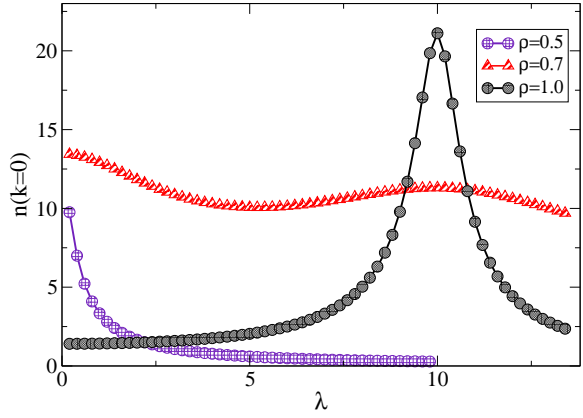}
\caption{(Color online) $n(k=0)$ against various values of $\lambda$
for densities, $\rho=0.5,0.7$ and 1.0.} \label{fig:fig6}
\end{figure}
%\vspace{2.0cm}
\begin{figure}[bp]
 \centering
\epsfig{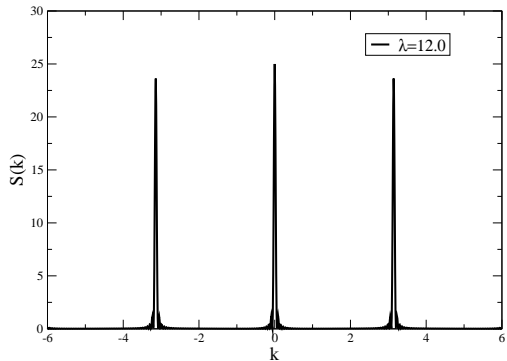}
\caption{Structure function, $S(k)$ versus the momentum, $k$, for
density $0.5$. } \label{fig:fig7}
\end{figure}

The momentum distribution is then calculated using the Eq.(\ref{eq:momdistri}). The value of $n(k)$ at $k=0$ is plotted against $\lambda$ for three
different values of $\rho$ ($0.5$, $0.7$ and $1.0$). A finite value of $n(k)$ at $k=0$ is a signature of the SF phase in the system.
These plots are
compared with  the chemical potential versus $\rho$ plots,
Fig.~(\ref{fig:fig2}), to confirm the gapped and gapless phases. In
the Fig.~(\ref{fig:fig6}), for $\rho=0.5$,  $n(k=0)$ has a very high
value when $\lambda$ is smaller than $0.8$, which signifies the SF phase. As
$\lambda$ increases beyond $0.8$, $n(k=0)$ falls off to values very close to
zero, implying certain phase transition from the SF phase to an insulating phase. This insulating phase is the SLMI-I as confirmed from the $\langle{\hat{n}_i}\rangle$
 plots [Fig.(\ref{fig:fig4})]. For $\rho=1$, $n(k=0)$ starts with
values very close to zero [Fig.(\ref{fig:fig6})]. Due to the very high value of $U$, the
system resides in the MI state. As $\lambda$ becomes comparable to
$U$, $n(k=0)$ attains a finite peak value, thus implying the
transition from the MI to SF phase. This result is in agreement with the ones obtained from the chemical potential plots [Figs.(\ref{fig:fig2}) and (\ref{fig:fig3})].
As $\lambda$ is increased further, $n(k=0)$ again drops to values very close to zero implying another
phase transition, but this time the transition being from SF to SLMI-II. Hence it can be said that this SF
is not the usual SF which comes due to $t$/$U$ ratio, but a consequence of the competition between
 the parameter $U$ and $\lambda$. At the incommensurate densities, for example, $\rho=0.7$ [Fig(\ref{fig:fig6})],
 it is observed that $n(k=0)$ stays finite for all values of $\lambda$, without any
signs of it vanishing, showing the presence of SF phase throughout the values of $\lambda$ considered in this work.
%\vspace{0.5cm}

\begin{figure}[tbp]
\centering
 \epsfig{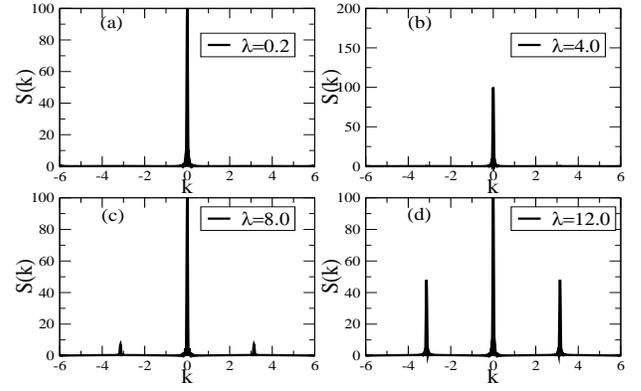}
\caption{Structure function, $S(k)$ versus the momentum, $k$, for density 1.0.}
\label{fig:fig8}
\end{figure}

\begin{figure}[tbp]
 \centering
\epsfig{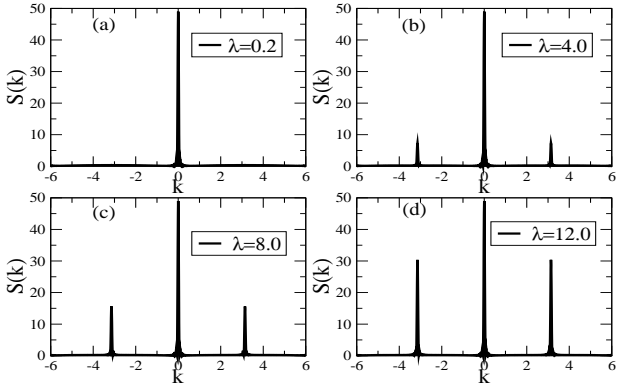}
\caption{Structure function, $S(k)$ vs the momentum, $k$, for density 0.7 for different values of $\lambda$.}
\label{fig:fig9}
\end{figure}

Figs.~(\ref{fig:fig7}~-~\ref{fig:fig9}) gives the structure function
as a function of $k$ for various values of density ($\rho=0.5$,
$1.0$ and $0.7$, respectively). In the normal optical lattice, the
periodicity is one lattice site. So for a MI phase, the structure
function peaks at $k=\pm 2\pi$~\cite{scarola}. However in an optical
superlattice, the periodicity is doubled. As a result, the
 Brillouin zone is halved. Hence in the MI phase for an optical superlattice, we expect to observe the peaks of $S(k)$ at $k=\pm\pi$.
 It can also be observed that for $\rho=1$, the peaks do not appear for
lower values of $\lambda$ [Fig.~(\ref{fig:fig8})]. The onset of the peaks actually signify the
effects of the superlattice potential, $\lambda$, in the system~\cite{rousseau}, and thus
changing the Brillouin zone boundaries. From Fig.~(\ref{fig:fig7}), we see two well defined peaks at $k=\pm \pi$ for $\rho=0.5$ at a very large value of $\lambda$ implying the presence of SLMI-I phase.
At $\rho=0.7$, we see two peaks at $k=\pm \pi$, for $\lambda$ more than $\approx 2.0$. This is a consequence of some density oscillations induced in the system by the superlattice potential.
%\vspace{0.5cm}

%\vspace{2.0cm}

\begin{figure}[bp]
 \centering
\psfig{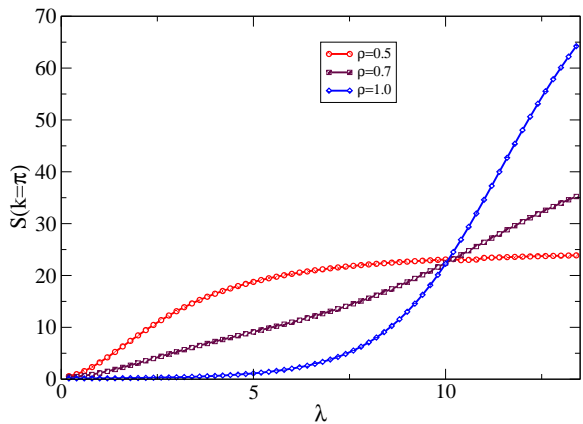}
\caption{(Color online) S($k=\pi$) versus $\lambda$ for three different density values, $\rho=0.5,0.7$ and $1.0$.}
\label{fig:fig10}
\end{figure}

The value of structure function at $k=\pi$ is plotted in
Fig.~(\ref{fig:fig10}) for some fixed densities against $\lambda$
to see how the effect of superlattice potential evolves in the
system. For $\rho=0.5$, $S(\pi)$ starts from very close to zero,
increases steadily and then becomes almost constant. For $\rho=1$,
initially there is no peak in the $S(k)$ plot at $k=\pi$. As
$\lambda$ is increased, even in the MI phase, we see a
finite value of the structure function at $k=\pi$, indicating the
onset of the effect of the superlattice potential. Once the
$\lambda$ overcomes $U$, the value of $S(\pi)$ increases rapidly,
showing the onset of the SLMI-II phase. It is also noted that at $\lambda\approx 10$, $S(k=\pi)$ is finite, showing some density modulations in the system. As seen earlier, at the above value of $\lambda$, the system undergoes a transition from MI to SF phase. Thus the SF at $\rho=1.0$ coexists with a finite density oscilallation in the system. For $\rho=0.7$, $S(\pi)$ is
very close to zero for low values of $\lambda$, and then increases
as $\lambda$ is greater than some value, and then goes on increasing
implying the setting in of some density modulations in the system. Hence finite peaks of $S(k)$ at $k=\pi$
[Fig.~(\ref{fig:fig9})] for $\rho=0.7$ implies the effect of
superlattice potential in the system bringing about density
modulation, although the system is in superfluid phase as indicated
from $n(k=0)$ plot [Fig.~(\ref{fig:fig6})] for all values of $\lambda$. So we see a coexistence of SF and some sort of
density oscillations in the system at the incommensurate density of $0.7$.

\begin{figure}[htbp]
\centering
\psfig{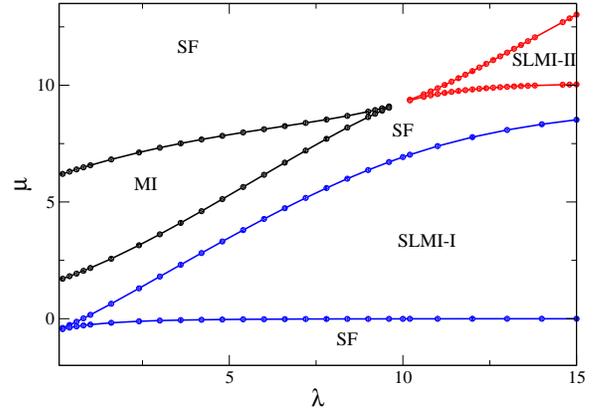}
\caption{(Color online) Phase diagram for the system of optical superlattice.}
\label{fig:fig11}
\end{figure}

The phase diagram in the $\mu$ - $\lambda$ plane for $U=10$ is
presented in Fig.~(\ref{fig:fig11}). The phase diagram consists of SF, MI,
SLMI-I and SLMI-II phases depending on the density and the superlattice potential.
\vspace{0.5cm}

\vspace{2.0cm}

\section{CONCLUSION}
We have analyzed the various phases exhibited by a system of bosons
in an optical superlattice with an unit cell consisting of two lattice sites, using the FS-DMRG method for
a large value of the on-site interaction, $U$. We find that at
density, $\rho=1/2$, the system is initially in the gapless
superfluid phase. But as the value of the superlattice potential,
$\lambda$ is increased, a finite gap arises, which corresponds to
the SLMI-I phase corresponding to the configuration [1 0 1 0 ...].
At integer density, $\rho=1$, the system is initially in the gapped
MI phase because of the large value of $U$. But as the superlattice
potential value is increased, the gap keeps on decreasing, and
ultimately shrinks to zero when $\lambda{\approx}U$. This gapless SF phase coexists with a finite $S(k=\pi)$ in the system implying density oscillations.
 As the value of
$\lambda$ is further increased, the gap reopens, corresponding to
the SLMI-II phase, with a configuration [2 0 2 0 ...]. At the incommensurate density,
$\rho=0.7$, we find the simultaneous existence of the SF phase along with density oscillations in the system.

\section{Acknowledgement} R.~V.~P. acknowledges financial support from CSIR and
DST, India. We are grateful to William Phillips,Yoshiro Takahashi, Mikhail Baranov,
Gora Shlyapnikov and Arun Paramekanti for stimulating discussions.

\end{document}